\documentclass[twocolumn,preprintnumbers,amsmath,amssymb]{revtex4-2}
\usepackage{graphicx}
\usepackage{color}
\usepackage{caption}
\usepackage{subfigure}
\begin{document}

\title{A water structure indicator suitable for generic contexts: two-liquid behavior at hydration and nanoconfinement conditions and a molecular approach to hydrophobicity and wetting}

\author{Nicol\'as A. Loubet$^1$}
\author{Alejandro R. Verde$^1$}
\author{Gustavo A. Appignanesi$^{1 *}$}

\newcommand{\angstrom}{\mbox{\normalfont\AA}}

\affiliation{
$^1$ INQUISUR, Departamento de Qu\'{i}mica, Universidad Nacional del Sur (UNS)-CONICET, Avenida Alem 1253, 8000 Bah\'{i}a Blanca, Argentina\\
* Corresponding author: appignan@criba.edu.ar\\
}

\date{\today}

\begin{abstract}

In a recent work we have briefly introduced a new structural index for water that, unlike previous indicators, was devised specifically for generic contexts beyond bulk conditions, making it suitable for hydration and nanoconfinement settings. In this work we shall study this metric in detail, demonstrating its ability to reveal the existence of a fine-tuned interplay between local structure and energetics in liquid water. This molecular principle enables the establishment of an extended hydrogen bond network, while simultaneously allowing for the existence of network defects by compensating for uncoordinated sites. By studying different water models and different temperatures encompassing both the normal liquid and the supercooled regime, this molecular mechanism will be shown to underlie the two-state behavior of bulk water. Additionally, by studying functionalized self-assembled monolayers and diverse graphene-like surfaces, we shall show that this principle is also operative at hydration and nanoconfinement conditions, thus generalizing the validity of the two-liquids scenario of water to these contexts. This approach will allow us to define conditions for wettability, providing an accurate measure of hydrophobicity and a reliable predictor of filling and drying transitions. Hence, it might open the possibility of elucidating the active role of water in broad fields of biophysics and materials science. As a preliminary step, we shall study the hydration structure and hydrophilicity of graphene-like systems (parallel graphene sheets and carbon nanotubes) as a function of the confinement dimensionality.  

FOR A PROGRAMMING CODE TO IMPLEMENT V4s, PLEASE GO TO: https://github.com/nicolas-loubet/V4S

\end{abstract}

%\pacs{83.80.Iz, 47.57.Qk, 83.85.Ei}

\maketitle

\section{Introduction}

Contrary to the simplicity that emanates from the austere beauty of its molecular formula, water is a system characterized by an extremely complex behavior\cite{chaplin,water_chemrev, ball,water1_epje,water2_epje}. In fact, water is known to present a plethora of structural, thermodynamic and dynamical anomalies\cite{chaplin,water_chemrev, ball,water1_epje,water2_epje}. And these anomalies are responsible for making it irreplaceable in relevant hydration and nanoconfinemt contexts\cite{chaplin,water_chemrev, ball,water1_epje,water2_epje,debenedetti,tanaka, protein,protein1,subnano,COMMENT-PRL,cavities,protein2,protein3,PRE-protein,graphene,membrane,martelli_franzese}. But how is it possible to bring in this level of complexity to a system represented by such a simple triatomic molecule? It seems that Nature has accomplished it by furnishing water with peculiar intermolecular interactions\cite{chaplin,water_chemrev, ball,water1_epje,water2_epje,v4,v4s}. Indeed, water presents hydrogen bonds (HB) that are stronger than other intermolecular forces typical of simple liquids and are endowed with directional nature, thus promoting the adoption of an open or expanded structure. Hence, contrary to simple liquids whose molecules compact their structure in order to enhance the number and strength of their intermolecular interactions, water molecules present the atypical capability of expanding their local structure (expanding their second molecular shell) in order to lower energy by improving the HBs with their (only) four tetrahedrally-arranged first neighbors, so that energy decrease goes hand in hand with volume expansion. In this sense, water displays a prominent tendency to maximize four-fold coordination to build up an extended HB network, with a small fraction of undercoordination defects whose population increases very moderately with temperature\cite{water_chemrev,water1_epje,water2_epje,v4s}. But the first-shell energetic improvement by second-shell volume expansion seems to be only part of the story. Creation of undercoordination defects within such an open-like HB network implies a significant coordination-energy loss. Thus, an additional structural plasticity is in order so that local contraction and reorientation of the second shell can provide the water molecule with partial energetic compensation to the lacking first-shell HB\cite{v4s}. This overall structural plasticity, embodied in a molecular principle based on this structure-energy mutual interplay, materializes in two local molecular arrangements, with low and high local density, as we have recently shown\cite{v4s}. These two molecular arrangements, in turn, lie at heart of the two-liquids scenario of liquid water, the seminal description that long ago shed light upon  this field\cite{criticalpoint,water_chemrev,water1_epje,water2_epje,v4s}. This description emerged from the discovery of the existence of a second liquid-liquid critical point in the supercooled regime\cite{criticalpoint,water_chemrev,water1_epje,water2_epje,LLCP_T1,LLCP_T2,LLCP_T3,LLCP_E1,LLCP_E2,LLCP_E3,Kim,Biddle,francesco-pablo}, implying the coexistence of two competing phases with different density (a supercooled liquid arises by cooling down below the melting point fast enough to avoid crystallization towards the glass state\cite{appignanesi06,dclusterswater,appignanesi07,link}). In fact, it implies an extension into the (supercooled) liquid regime of the coexistence line between the low density amorphous ice (LDA) and the high density amorphous ice (HDA)\cite{criticalpoint}. But while the structure of the low local density phase has been long recognized, the structure of the high local density one has been largely overlooked even when, contrary to the situation in simple liquids, there is a significant energetic penalty associated with the creation of defects within the HB network of water which, as already indicated, would demand local structuring to bring partial compensation\cite{v4s}. In turn, the two-liquids scenario has motivated the development of several structural indicators along the years to characterize the two competing local molecular arrangements in liquid water\cite{Errington,d5,LSI,LSIGAA,zeta,LOM,Foffi,EPJPlus,v4,v4s,v4T2}. However, these indices have been devised for bulk water and might present artifacts when applied to more general contexts like that of hydration and nanoconfinement\cite{COMMENT-PRL}. In a recent work to investigate the behavior of TIP3P water at ambient temperature for non-bulk conditions, we preliminarily introduced  a new metric inspired by the directional nature of the HB interactions (which we called $V_{4S}$), suited for generic contexts\cite{v4s}. Thus, in the present work we shall study in detail the behavior of this indicator for different water models (with special emphasis on TIP4P/2005) and for different temperatures covering from ambient to the supercooled regime. We shall also apply it for certain hydration and nanoconfinement settings (specially for graphene-like systems) in order to define conditions for wettability. In so doing, we shall also evidence its potential to measure and rationalize hydrophobicity from a molecular level.    

\section{The V$_{4S}$ structural indicator for water}
%\subsection{The V$_4s$ structural indicator for water}

\begin{figure}[h!]
\resizebox{0.5\textwidth}{!}{%
\includegraphics{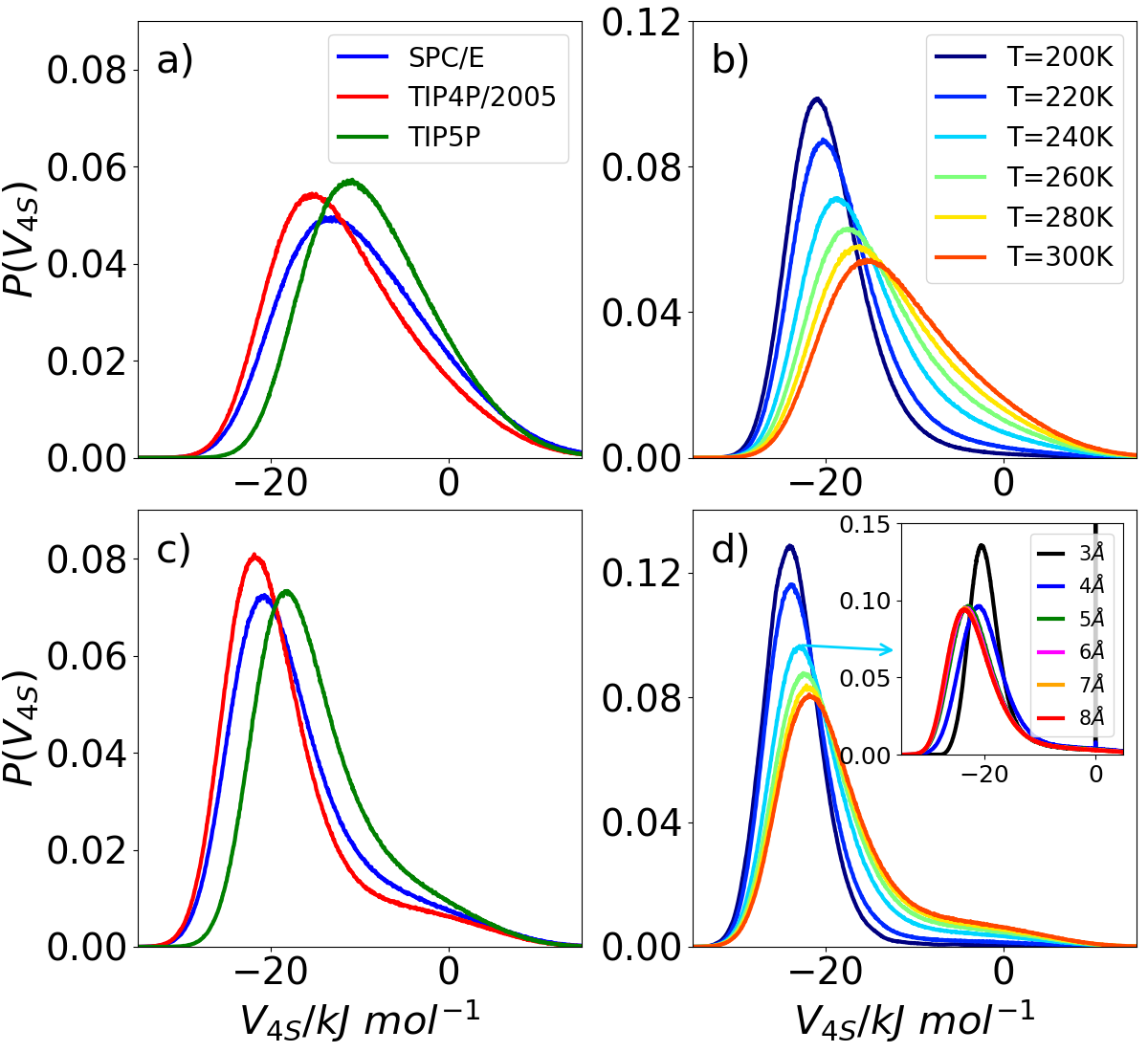}
}
\caption{Distributions of the $V_{4S}$ index for bulk water. We show different water models at T=300K and 1 bar and we also display the behavior of the index for TIP4P/2005 at different temperatures, encompassing the liquid and supercooled regimes. a) and b) use the real dynamics scheme (RD, instantaneous molecular dynamics configurations), while c) and d) present the corresponding plots at the inherent structures scheme (IS, configurations subject to potential energy minizations). The inset of Fig.~\ref{fig1} d) displays the $V_{4S}$ distributions that would arise for different values of the cutoff $R$ ranging from 3 to 8\AA\ for the case of TIP4P/2005 water model at T=240K.}
\label{fig1}
\end{figure}

In a recent work we preliminarily introduced a new parameter to characterize water structure in terms of its directional (tetrahedral) intermolecular interactions. It was named $V_{4S}$ since, for any central water molecule, it computes the different energetic contributions at four tetrahedrally-arranged sites. Unlike other structural indicators that might produce artifacts at non-bulk conditions (like at hydration or under nanoconfinement)\cite{COMMENT-PRL}, $V_{4S}$ was specially devised to be suitable for generic contexts, since the energy contributions at each of the four tetrahedral sites characteristic of a water molecule could originate from the interaction with any kind of neighboring heavy atoms besides the oxygens of other water molecules. Specifically, for any given central water molecule we first open its H-O-H angle to reach the value of a tetrahedral angle ($arccos(-1/3)$, unless the H-O-H angle is already tetrahedral, as in the case of the SPC/E model). Then, we place two tetrahedral point/sites 1\AA\ away from the oxygen along these redefined O-H lines and build a perfect tetrahedron centered at the oxygen that includes these sites as two of its vertices. In this sense, we are left with four tetrahedrally-arranged points/sites (the vertices of the tetrahedron) which partition space in four regions, since we assign all the neighboring heavy atoms located at less than $R = 5\angstrom$ from each of the four sites (be they oxygens from neighboring water molecules or the heavy atoms of any other neighboring system) to their closest tetrahedral site. Subsequently, for each tetrahedral site we add up the contributions of all the pair-wise interaction potentials between the oxygen of the central water molecule and all its assigned heavy atoms, including both Lennard-Jones and Coulomb interactions. Hence, we are left with four potential values, one for each point/site of the tetrahedron, which we finally order from $V_{1S}$ (the most interacting, that is, the lowest potential value) to $V_{4S}$ (the least interacting or highest value). The use of a cutoff of $R = 5\angstrom$ from each tetrahedral site is justified since this is the distance at which the resulting value of the $V_{4S}$ begins to converge. As an example, in Fig.~\ref{fig1} we show results for the index distribution when calculated with a cutoff $R$ from 3 to 8\AA\ for TIP4P/2005 water model at T=240K.

\begin{figure}[th]
% \begin{center}
%\mbox{
		%\leavevmode
     %\subfigure 
				{\includegraphics[width=8.5cm]{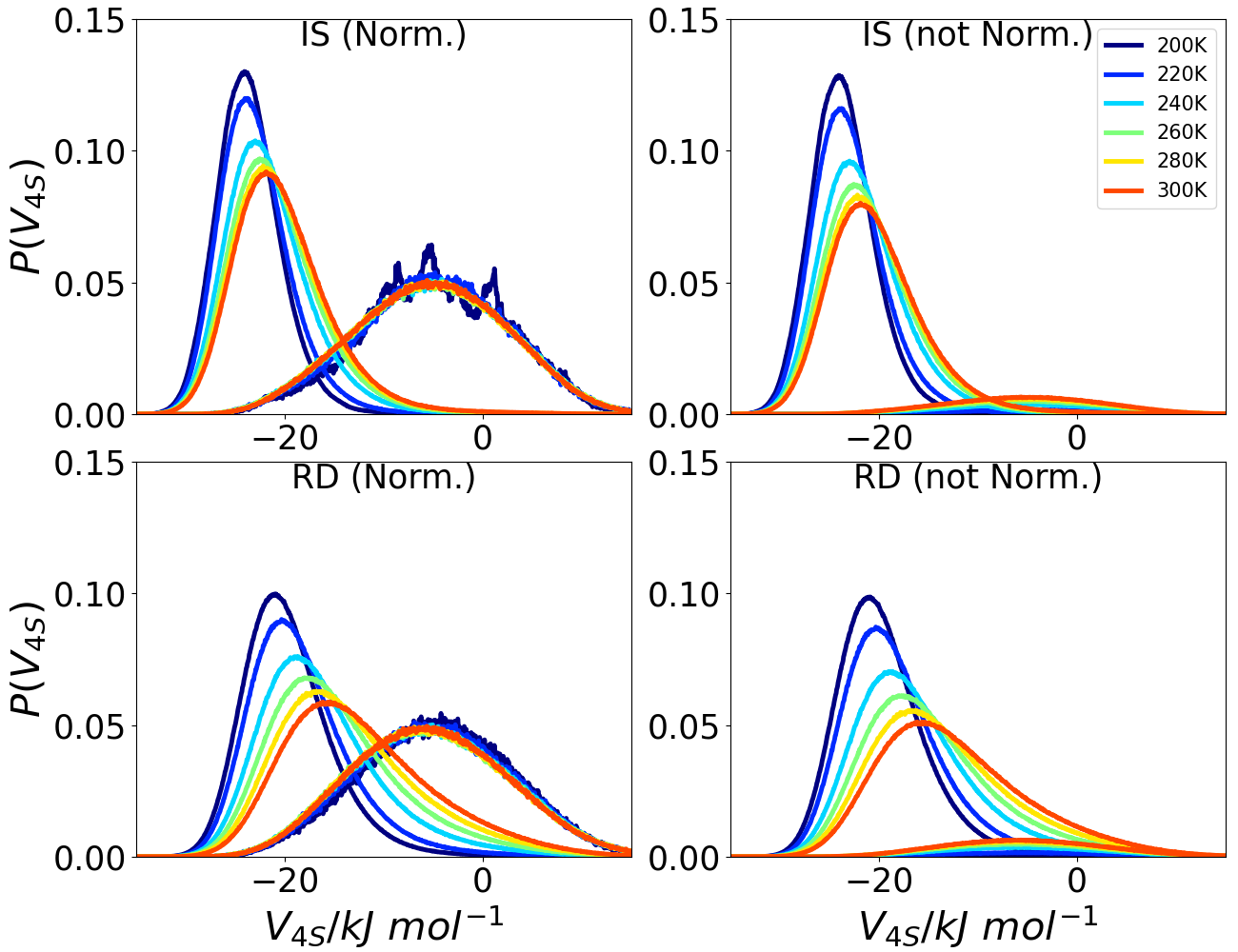}}
 %				{\includegraphics[width=8.5cm]{fig1_b}}
%				{\includegraphics[width=8.5cm]{zetaspce}}
%				}
%	\end{center}
	%\vskip -0.7cm
\caption{Distributions of the  $V_{4S}$ for T molecules (with four hydrogen bonds) and D molecules (three hydrogen bonds) for TIP4P/2005 water at different temperatures. The plots on the left side show the case when all curves are normalized to display unitary area, while in the ones on the right they are affected by the fraction of the corresponding kind of molecules. The plots at the top side correspond to the inherent structures scheme while the ones at the bottom are for the real dynamics. 
}
\label{fig2}
\end{figure}

\begin{figure}[th]
% \begin{center}
%\mbox{
		%\leavevmode
     %\subfigure 
				{\includegraphics[width=8.5cm]{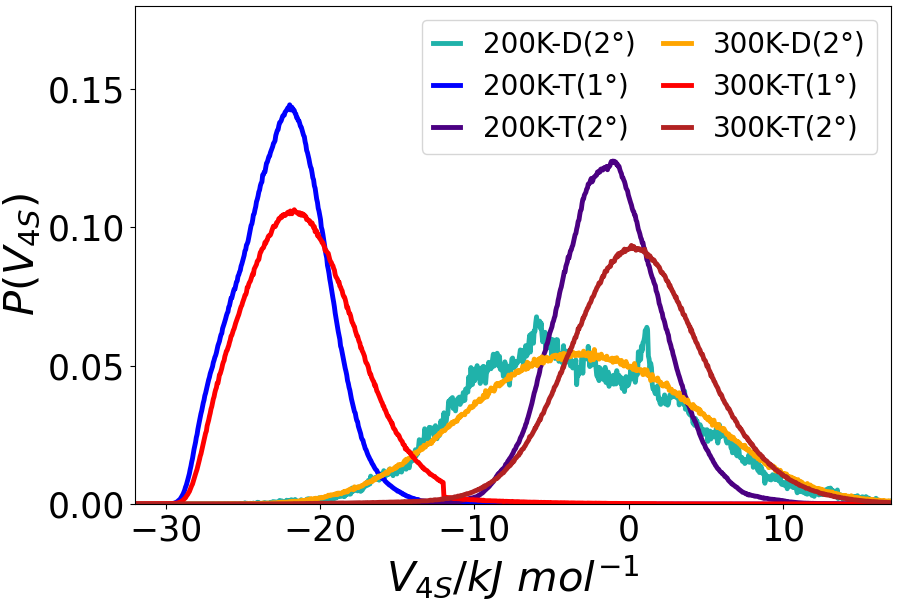}}
 %				{\includegraphics[width=8.5cm]{fig1_b}}
%				{\includegraphics[width=8.5cm]{zetaspce}}
%				}
%	\end{center}
	%\vskip -0.7cm
\caption{Contributions to the $V_{4S}$ of the first ($1^\circ$) and second ($2^\circ$) neighbor shells of T molecules at two different temperatures. For the D molecules we only show the second shell data since they only present three neighbors in the first shell. The small step in the curve at T=300K for the first neighbor shell contribution is due to the threshold used to classify molecules as T or D (note that it occurs precisely at a value of $-12 kJ/mol$)
}
\label{fig3}
\end{figure}

Fig.~\ref{fig1} shows the distribution of the $V_{4S}$ index for different water models at ambient conditions. In the case of TIP4P/2005, we also consider different temperatures from above to below its melting point, estimated to be at around T=250K \cite{cvega} (that is, including both the liquid and the supercooled regime). We show results for the instantaneous or real dynamics (RD, configurations of the molecular dynamics trajectory) and for the inherent dynamics (inherent structures, IS; configurations subject to potential energy minimization by using the steepest-descent method in order to get rid of the blurring effect of thermal energy\cite{v4,v4T2}). From direct inspection of such plots clear signs of bimodality are evident, mainly at the inherent dynamics, which is the scheme we shall use from now on. This fact is even more evident when we discriminate the water molecules in tetrahedral (T) or defective (D) as done by the previously introduced $V_{4}$ indicator (suitable only for bulk water) that finds the four water molecules with which a given molecule interacts more strongly and picks the value of the less intense of such interactions\cite{v4,v4T2}. In simple terms, the T molecules (with low local density) are the ones with $V_{4} \le -12 kJ/mol$, which basically means that they present four good tetrahedrally-arranged hydrogen bonds with their first neighbors (we consider as HBs the interactions stronger than $-12 kJ/mol$ at the inherent dynamics, a criterion that has been shown to give results consistent with geometric considerations, as we have already noted\cite{v4}). In turn, the D ones (with high local density) represent undercoordination defects since they present only three HBs (that is, $V_{4} > -12 kJ/mol$)\cite{v4,v4T2}. The T molecules are majority at all temperatures studied, while the D ones are extremely scarce at low supercooled temperatures and their population grows moderately as temperature increases. Fig.~\ref{fig2} shows the relative contributions of the T and D molecules to the $V_{4S}$ index. While the former are responsible for the presence of the sharp peak at the left, the latter contribute to a  broader peak located at the right hand side. We show normalized distributions (with curves for the T and D molecules both presenting unitary areas) and we also present the distributions affected by their relative populations (so that the sum of the areas under the T and the minority D molecules add to unity). We depict results both for the real and the inherent dynamics, but from now on we shall use exclusively the inherent dynamics approach. While the peak of the T molecules moves slightly with temperature and with the minimization (the structure improves as thermal energy is removed), the peak for the D molecules is insensitive to such changes. For the T molecules, expansion of the second molecular shell allows the first shell coordination to improve. Thus, the contribution to $V_{4S}$ comes mainly from first shell energetics (a good quality HB) while the farthest second shell does not contribute significantly. On the contrary, in the case of the D molecules, there is a lacking HB, so there is basically no contribution to $V_{4S}$ from the first shell and all the contribution comes from the second shell which contracts and orients in order to compensate such a great energetic loss\cite{v4s}. Indeed, Fig.~\ref{fig2} tells us that the second shell contribution for central D molecules is significant, with a mean value of around $-6 kJ/mol$. These facts are more evident from Fig.~\ref{fig3} where we show the discrimination between the contributions to $V_{4S}$ of the first and second neighbor shells for the T molecules and the contribution of the second neighbor shell for the D ones. We use a simple geometric criterion of 3.2\AA\ to separate between the two neighbor shells which is quite effective at the inherent dynamics, since the region around such value is depopulated by the minimization scheme that either restores deformed hydrogen bonds or moves away intershell molecules to their typical interstitial positions at around 
3.5\AA \cite{HDL}. This again makes evident the advantage of the inherent dynamics approach for classification purposes. While the second shell for the T molecules shows distributions with mean value close to zero and, thus, the corresponding $V_{4S}$ values are almost exclusively due to the energetics of the first shell, the contribution of the second shell for the D molecules involves significant negative (attractive) values. Hence, the existence of a fine-tuned interplay between structure and energy is evident:
For the LDA-like T molecules there is an expansion of the second shell that enables a first shell energetic improvement so that the net attraction is (depending on temperature) close to $-20 kJ/mol$. In turn, for the D molecules (the ones involved in the HDA-like state) there is a contraction of the second shell at the lacking HB-site that indeed promotes a partial energy compensation of the HB loss, since we find that the mean value is quite attractive (around $-6 kJ/mol$, with a few second shell molecules involved in this energetic compensation). These results lead to a relevant observation: the use of $V_{4S}$ unravels the molecular principle that underlines the two-state picture for water, revealing a marked structural plasticity to enable the energetic requirements for the existence of two different local molecular arrangements. Thus, this description encompasses the existence of an extended HB-network of tetrahedrally-coordinated molecules that allows for the occurrence of hydrogen-bond undercoordination defects in a structure-energy compensation scheme involving the two first coordination shells.

\begin{figure}[th]
% \begin{center}
%\mbox{
		%\leavevmode
     %\subfigure 
				{\includegraphics[width=8.5cm]{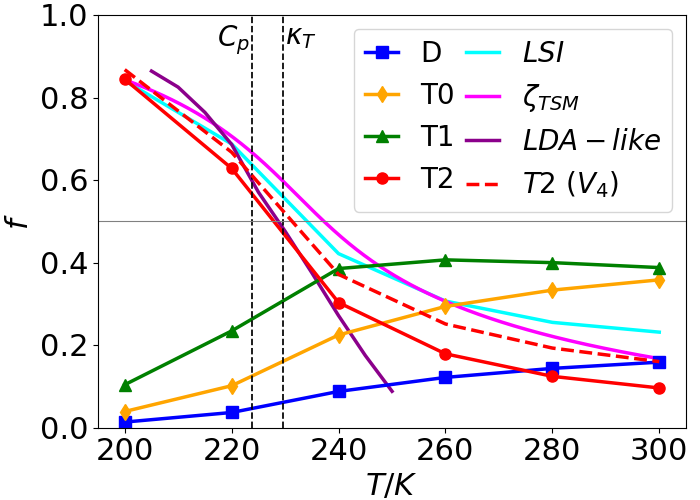}}
 %				{\includegraphics[width=8.5cm]{fig1_b}}
%				{\includegraphics[width=8.5cm]{zetaspce}}
%				}
%	\end{center}
	%\vskip -0.7cm
\caption{Fractions of the different molecular classes as obtained with the $V_{4S}$ index (D, T0, T1 and T2) as a function of temperature. The T2 ones represent the structured LDA-like state while the rest of the molecules would conform the HDL-like one. We also include the fractions of the structured state as estimated by other indicators like the local structure index (LSI) at the inherent structures\cite{LSIGAA}, the $V_4$ parameter (the fraction of T2 molecules calculated with this index)\cite{v4T2}, the $\zeta$ index\cite{zeta} and the LDA-like from a Neural Network approach\cite{Fausto-Francesco-ML}. The vertical dashed lines indicate the temperatures of the $C_p$ and $K_T$ maxima. The values for the $\zeta$ index and for the LDA-like state of the Neural Network approach were taken from \cite{Fausto-Francesco-ML}
}
\label{fig4}
\end{figure}

Since at all temperatures studied (even above the melting point) the system consists basically of a sea of T molecules with a few defects (D molecules) whose population grows moderately with increasing temperature, a description of the two-liquid scenario on the basis of a single-molecule approach does not seem reasonable and must be exceeded towards a multi-molecular scheme\cite{v4,v4T2}. Indeed, it is obvious that while the low density phase of water might still be considered as a pure T-state, the high density phase should be represented by a mixed state given the fact that the D molecules are mainly surrounded by T molecules. In a previous work, by using the $V_4$ indicator\cite{v4T2}, we already classified molecules incorporating their neighborhood. The relevance of the incorporation of the local neighborhood had also been already considered as in the introduction of the local order metric\cite{LOM}, in the coarse-grained evolution of the $\zeta$ indicator\cite{zeta} and in studies of water defects\cite{bagchi}. Thus, we shall now incorporate information from the local molecular neighborhood to the $V_{4S}$ index. First of all, we shall classify water molecules as T and D but now directly using the $V_{4S}$ instead of the $V_4$ index as done above. We use the same energetic threshold as done for the $V_4$ index, $V_{4S}=-12kJ/mol$. Values lower than this threshold (more attractive) indicate that the fourth tetrahedral site (the least interacting of the four tetrahedral sites of the water molecule) feels an attraction compatible with a good quality hydrogen bond. Higher values imply that such tetrahedral site lacks a good coordination energy and should be considered as a defect. We note that we obviously expect that this classification with the $V_{4S}$ index would yield results similar to that of the $V_4$ one for bulk water. The clear advantage of the $V_{4S}$ is that it is valid beyond bulk water, being suitable for conditions of hydration and nanoconfinement. Then, having dissected the water molecules into T and D ones, we now further classify the T ones into classes T0, T1 and T2: For a T molecule, if the interaction at one of the first four tetrahedral sites is contributed by a D molecule, we classify such T molecule as a T0 one. Then for each of the remaining T molecules, if the contribution at one of its four tetrahedral sites is provided by a T0 molecule, we classify them as T1. Finally, all remaining T molecules are classified as T2 ones. This classification scheme means that T0 molecules present a D molecule within their first coordination shell, that T1 molecules present D ones as second coordination neighbors and that the two first coordination shells of the T2 molecules are free of molecular defects. In Fig.~\ref{fig4} we show the temperature dependence of the populations of D, T0, T1 and T2 molecules. The population of T2 molecules, which is expected to represent the low-density state, is high at low temperatures where there is a very scarce fraction of D molecules, but declines continuously as temperature increases. The isofraction between the low density-like state (T2) and high density-like one (sum of D, T0 and T1) together with the inflection point in the T2 curve occur in the region around the maxima of the isothermal compressibility ($K_T$)\cite{Biddle} and the isobaric heat capacity ($C_p$)\cite{Biddle,v4}, which is consistent with the high fluctuations that are expected to emerge from the competition of the two structural states having equal probability, as prescribed by the two-state model.

\section{V$_{4s}$ at hydration and nanoconfinement conditions. A molecular approach to hydrophobicity and wetting}

In this section we shall focus on model hydration and nanoconfinement settings. We shall make use of self-assembled monolayers, SAMs, which consist of arrays of alkyl chains functionalized by ending with hydrophilic ($OH$) or hydrophobic ($CH_3$) groups\cite{review_Garde,v4s}, graphene sheets and narrow single-walled carbon nanotubes\cite{v4s,graphene,cavities,Fluid-Phase-Equilibria,HummerCNT}. 

\begin{figure}[th]
% \begin{center}
%\mbox{
		%\leavevmode
     %\subfigure 
				{\includegraphics[width=8.5cm]{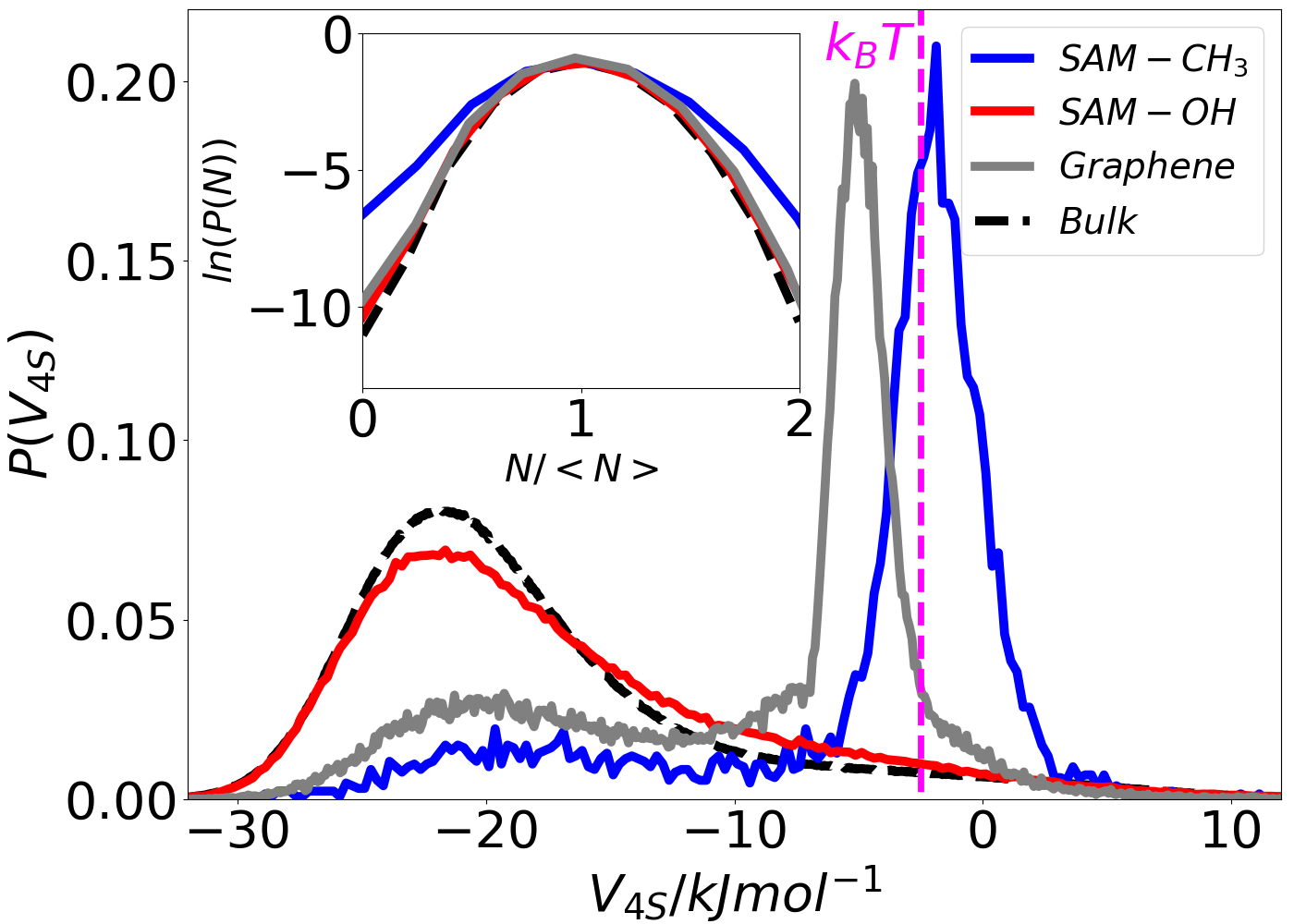}}
 %				{\includegraphics[width=8.5cm]{fig1_b}}
%				{\includegraphics[width=8.5cm]{zetaspce}}
%				}
%	\end{center}
	%\vskip -0.7cm
\caption{Distributions of the $V_{4S}$ index for a hydrophobic SAM ($SAM-CH_3$), a hydrophilic SAM ($SAM-OH$) and a graphene sheet. In all cases, the systems are solvated with TIP4P/2005 water molecules at a temperature of 300K (we also include the corresponding $V_{4S}$ distribution of bulk water). Inset: Probability distributions for observing N heavy atoms, P(N), within a small spherical observation volume of 3.3\AA\ tangent to each surface and at bulk conditions.
}
\label{fig5}
\end{figure}

In Fig.~\ref{fig5} we present the distributions of the $V_{4S}$ index for the two self-assembled monolayers together with a graphene sheet at T=300K and 1 bar. We use the TIP4P/2005 water model as hydration water, and we evaluate water molecules that are closer than 3.3\AA\ from each surface (since water molecules can form HBs with the $SAM-OH$ surface and, thus, they can become closer to the surface as compared with the $SAM-CH_3$, so in this case we used a cutoff of 3.0\AA, but the results are robust if we increase the cutoff to a value similar to that of the $SAM-CH_3$). From Fig.~\ref{fig5} it is immediately evident that the $V_{4S}$ parameter clearly discriminates hydrophobic from hydrophilic behavior. The distribution for the hydrophilic SAM is very similar to that of bulk water, with a high peak consistent with LDA-like water molecules (below $-20kJ/mol$) and also evidencing population of HDA-like ones (at around $-6kJ/mol$). As expected, the OH moieties of the hydrophilic SAM replace one of the hydrogen bonds of the neighboring water molecules, thus offering them an environment similar to that of the LDA-like bulk molecules. Obviously, this is not the case for the hydrophobic SAM, whose $V_{4S}$ distribution shows a rather sharp peak very close to $-k_B T$ (where $k_B$ is Boltzmann constant and $T$ is the absolute temperature). Moreover, such peak has almost completely decayed at the region corresponding to the HDL-like peak of bulk water ($-6kJ/mol$). Since it is not possible to form a HB with the $SAM-CH_3$ surface, the neighboring water molecules lack coordination at their fourth tetrahedral site ($V_{4S}$). Even more, it is also evident that they are neither able to partially compensate this loss, as done by the bulk HDL-like molecules. In turn, the case of the graphene surface is very interesting since, while the water molecules cannot form a HB with the surface, the occurrence of a sharp peak close to $-6kJ/mol$, the same region of the HDL-like bulk water molecules, is evident. Also, the distribution has almost completely decayed at around $-k_B T$. This fact speaks of the occurrence of a certain hydrophilic behavior for graphene, as has been recognized both theoretically\cite{graphene,Fluid-Phase-Equilibria} and by careful experimental water contact angle measurements over clean graphene-based surfaces\cite{graphene-exp}. Hydrophobicity can be quantified by the extent of number density fluctuations of the local hydration water and, hence, by computing $P(N=0)$, the probability of emptying a small volume (like a sphere of radius $r=3.3\angstrom$ tangent to the surface), where N is the number of water molecules inside the local observation volume\cite{review_Garde}. This value is inversely proportional to the work of cavity creation at such site (a high/low probability value implies a low/high work of water removal) and, thus, provides an estimate of the local dehydration propensity. The value of $P(N=0)$ for the systems studied can be found from the corresponding ordinate values in the inset of Fig.~\ref{fig5}. Both the SAM-OH and bulk water present a very low (and very similar) value for water vacating probability. For the SAM-OH it implies that the water molecules are tightly hydrating the surface (they not only would display high residence times at the hydration layer, but it would be very difficult to remove water and create a cavity). On the contrary, the $SAM-CH_3$ exhibits a value that is two orders of magnitude larger, which is consistent with the fact that hydration water is readily removed from the hydration layer by thermal energy. Thus, it displays a large dehydration propensity provided the preference of the water molecules to be at bulk conditions (either as LDA-like of HDL-like states) since the surface is neither able to establish a HB nor to offer an environment with the partial energy compensation that provides the second shell of bulk water molecules at HB-lacking sites. In turn, the case of the graphene surface is quite appealing, since it displays a $P(N=0)$ value very close to that of the SAM-OH, thus implying that this surface behaves as quite hydrophilic. This extent of hydrophilicity (which means that the water molecules display a low dehydration propensity similar to surfaces capable of hydrogen-bonding water) is unexpected provided the fact that water molecules at graphene lack a HB in the direction of the surface and can only recover around $30\%$ of such energy loss by interacting with the (albeit dense) carbon network\cite{v4s,graphene,cavities,Fluid-Phase-Equilibria,HummerCNT}. Indeed, the peak of their $V_{4S}$ distribution occurs at values that are less than $1/3$ of the peak that the SAM-OH presents for the hydrogen bonded' molecules to the surface (located at values below $-20kJ/mol$). These strikingly contrasting observations can be reconciled if we consider that the graphene-like surface is able to offer the water molecules with an environment energetically similar to that of the bulk HDL-like molecules (close to $-6kJ/mol$). Thus, it is not necessary to recover full HB-like energetics, but it suffices to mimic the (partially compensating) energetics of the HDL-like molecules. This represents a relevant piece of information since it extends the validity of the two-state picture of water to non-bulk contexts. When full HB recovery is not possible, water molecules might need to partially compensate this lacking coordination site at the level of a HDL-like bulk molecule in order to provide a good hydrophilic wetting to the interacting surface. If the interaction decreases, being less intense than that corresponding to a HDL-like environment (that is, if the surface is not able to provide an environment that energetically mimics any of the two bulk-like states, either replacing a HB or partially compensating the HB-loss at least at the level done at bulk water), the level of hydrophilicity decays, until reaching the hydrophobic limit at a value of  $V_{4S}$ around $-k_B T$. At such hydrophobic conditions, a water molecule would be more comfortable at bulk contexts (being either a LDA-like or HDA-like molecule) than at the interface and, thus, the corresponding work for water removal (and, consequently, the mean interfacial residence time) would be low.

Having described the ability of the $V_{4S}$ index to successfully reveal the molecular basis of hydrophilic wettability, we now apply it to study nanoconfinement conditions. We shall consider water confined by graphene parallel planar plates and also small radius single-walled carbon nanotubes in order to deal with situation of different dimensionality from three to one dimensions. For the parallel graphene sheets we use two separations between the plates: 7.0\AA\ and 6.5\AA. While both cases are narrow, the latter is very close to the limit of separation that enables water filling since below such separation value the liquid state ceases to be thermodynamically stable within the inter-plate region and the plates collapse, as demonstrated by potential of mean force calculations\cite{graphene}. The lowest separation case, thus, basically consists in a two-dimensional confinement situation, while at the highest separation there is certain thickness available to begin to include more than a single water shell.

\begin{figure}[th]
% \begin{center}
%\mbox{
		%\leavevmode
     %\subfigure 
				{\includegraphics[width=8.5cm]{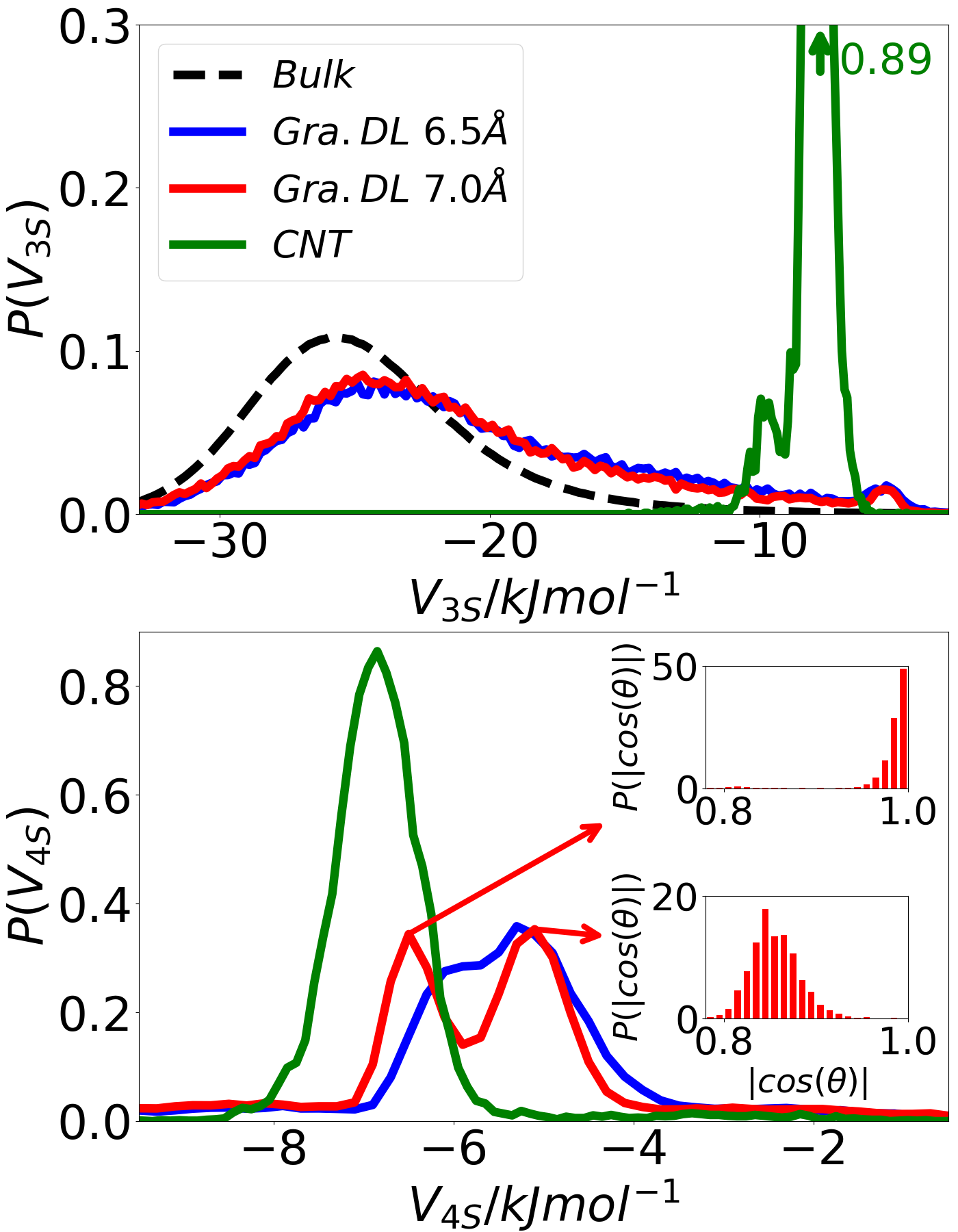}}
 %				{\includegraphics[width=8.5cm]{fig1_b}}
%				{\includegraphics[width=8.5cm]{zetaspce}}
%				}
%	\end{center}
	%\vskip -0.7cm
\caption{Distributions of $V_{4S}$ and $V_{3S}$ for the interior region of a pair of parallell graphene sheets (at two different separations, namely, $6.5\angstrom$ and $7.0\angstrom$). The insets depict the distribution of the orientational angle ($\theta$) of the virtual point $V_{4S}$ of the water molecules (considering the line from the oxygen to the least interacting tetrahedral site that defines the $V_{4S}$) with respect to the normal of the parallell graphene sheets. We depict the absolute value of the $\cos (\theta)$. We also include the distributions of $V_{4S}$ and $V_{3S}$ for the narrow carbon nanotube we studied.
}
\label{fig6}
\end{figure}

\begin{figure}[th]
% \begin{center}
%\mbox{
		%\leavevmode
     %\subfigure 
				{\includegraphics[width=8.5cm]{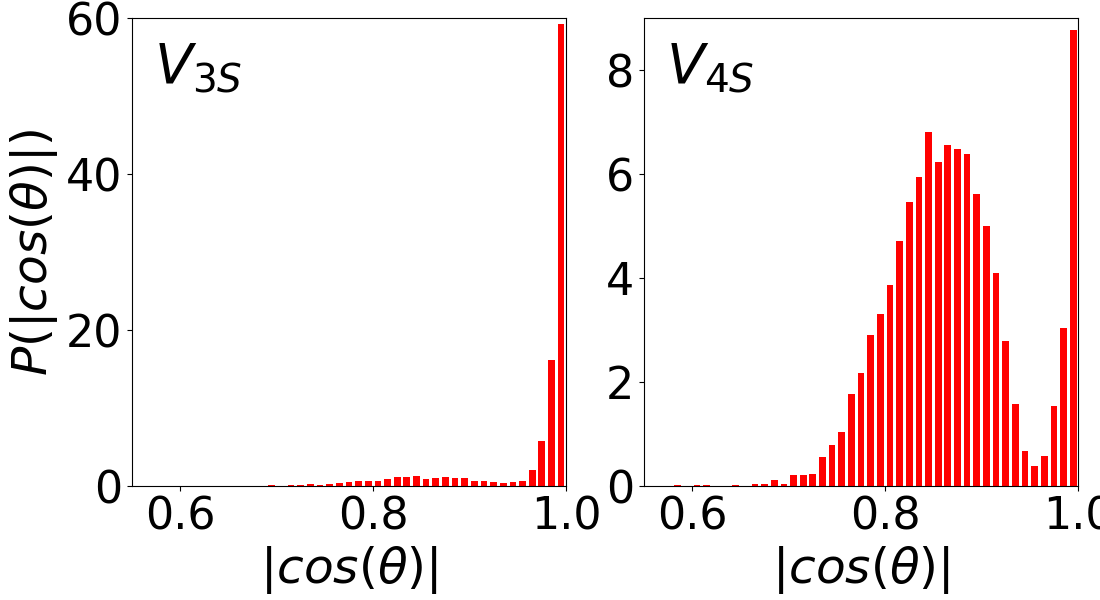}}
 %				{\includegraphics[width=8.5cm]{fig1_b}}
%				{\includegraphics[width=8.5cm]{zetaspce}}
%				}
%	\end{center}
	%\vskip -0.7cm
\caption{Orientational angle for the virtual points $V_{3S}$ and $V_{4S}$ of the water molecules with respect to the radial of the carbon nanotube.
}
\label{fig7}
\end{figure}

In Fig.~\ref{fig6} (bottom) we show the resulting $V_{4S}$ distributions for the systems under study. The peaks for the parallel graphene sheets occur at values close to $-6kJ/mol$, while the $V_{4S}$ peak in the case of the CNT is located a bit to the left. This is so since the CNT curvature enables the interior water molecules to interact with slightly more carbon atoms than in the case of a planar sheet. In the case of the parallel graphene sheets separated by the largest value (7.0\AA) there are two peaks for  $V_{4S}$. The one to the left corresponds to water molecules that maximize the interaction with one of the walls and, thus, the molecules acquire a normal position, orienting the tetrahedral site (corresponding to $V_{4S}$) to point directly to such graphene sheet (a $\theta$ angle of around $0^\circ$; see inset in Fig.~\ref{fig6}). The other peak, the one that occurs at a less negative interaction value (less attractive) includes water molecules whose orientation departs a bit from the normal one, with $\theta \approx 30^\circ$. When we consider the case of the parallel plates at the lowest separation, the peak at the right is conserved (corresponding to $\theta \approx 30^\circ$) but the one corresponding to normal orientation ($\theta \approx 0^\circ$) begins to disappear, moving to higher angle values towards $\theta = 30^\circ$. If the sterical constraint is incremented by moving the plates to closer separations, emptiness is promoted and the plates collapse to each other. In Fig.~\ref{fig6} we also depict the case for the interior of a small radius carbon nanotube, CNT. This setting constitutes a one dimensional confinement since the water molecules form a linear arrangement where each one is hydrogen bonded to other two water molecules. Here not only the tetrahedral sites corresponding to the $V_{4S}$ interacts with the carbon wall but also that of the $V_{3S}$ (since the water molecules form a one-dimensional arrangement in which each water molecule interacts with other two water molecules and both the other two tetrahedral sites interact with the CNT wall). The former peaks at around $-6.85 kJ/mol$ (see Fig.~\ref{fig6} (bottom)), while the latter shows a peak at an even more attractive value of around $-7.80 kJ/mol$ (Fig.~\ref{fig6} (top)). The orientations of the $V_{4S}$ with respect to the radial of the CNT occur at around $\theta \approx 30^\circ$ while the case of the $V_{3S}$ corresponds to normal orientations, that is, $\theta \approx 0^\circ$ indicating the improvement of the water-wall interactions (see Fig.~\ref{fig7}). From (Fig.~\ref{fig6} (top)), we can see that in the case of the parallel graphene sheets, their $V_{3S}$ distributions are both characterized by the presence of a peak consistent with a water-water HB, below $-20kJ/mol$, implying the development of a good quality HB network in the confinement region. In the case of the lowest plate separation, this peak is also present since it is at least possible to accommodate a two-dimensional water layer and, thus, each water molecule interacts with one plate via its tetrahedral site corresponding to $V_{4S}$, while the other three tetrahedral sites imply water-water interactions. If the separation of the plates is even lowered, dehydration and, thus, the collapse of the plates is promoted. 

Taken together, all these results for the SAMs and the graphene-like systems reinforce the notion of the tendency of the water molecules to retain directionality in their interactions. Four-fold tetrahedral orientation like that in bulk conditions is preferred even when some of such interactions are established with the hydrating surface, either by the occurrence of a water-surface HB or by partial directional energy compensation. However, when such contribution is below the one the water molecules would receive at a HDA-like bulk water environment, the surface is less able to retain the hydrating water molecules for long times, with the consequent increment in the surface hydrophobicity.  

\begin{figure}[th]
% \begin{center}
%\mbox{
		%\leavevmode
     %\subfigure 
				{\includegraphics[width=8.5cm]{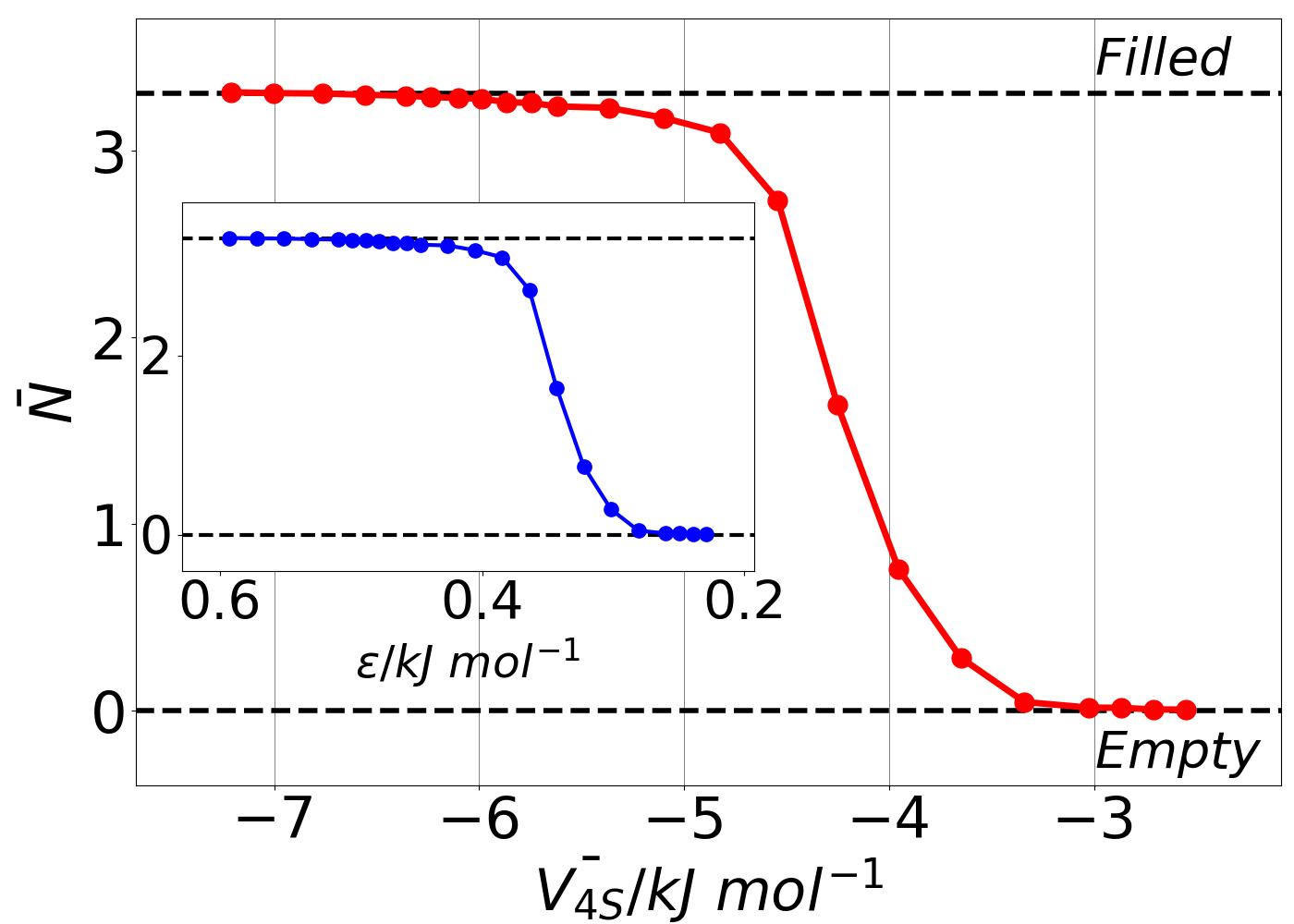}}
 %				{\includegraphics[width=8.5cm]{fig1_b}}
%				{\includegraphics[width=8.5cm]{zetaspce}}
%				}
%	\end{center}
	%\vskip -0.7cm
\caption{Average number of water molecules inside the CNT as a function of the average $V_{4S}$ value for CNTs with varying strenght of water-wall interactions. We include dashed lines that indicate the values for complete filling and emptiness. The inset shows the average number of water molecules as a function of the corresponding $\epsilon_{C-O}$ values characterising the water-wall attractions.
}
\label{fig8}
\end{figure}

For completeness, we wish to stress the ability of the $V_{4S}$ index to predict filling and drying transitions\cite{v4s}. The filling of subnanometric-radius carbon nanotubes like the one we have shown in Fig.~\ref{fig6} had been stated as ``surprising''\cite{HummerCNT,v4s} since only around $1/3$ of the HB coordination loss of the intruding water molecules is recovered by water-wall attractions, while partial reduction of such attractions is known to produce drying\cite{HummerCNT,v4s}. Thus, we now revisit the study of a narrow CNT we performed previously\cite{v4s}, which was hydrated by TIP3P water molecules, and we varied the Lennard-Jones parameters from $\epsilon_{C-O}=0.478369 \ kJ~{mol}^{-1}$ and $\sigma_{C-O}=3.27510$ \AA\ to $\epsilon_{C-O}=0.228720\ kJ~{mol}^{-1}$ and $\sigma_{C-O}=3.44154$ \AA. In Fig.~\ref{fig8} we provide the equilibrium average number of water molecules inside the CNT as a function of the $\epsilon$ parameter and of the corresponding $V_{4S}$ value (that is, the mean value of $V_{4S}$ for molecules inside the CNT when the attractions are varied). Depending on the conditions, we find an empty and a filled regime, separated by a transition one with partial occupancy (alternance of empty and filled periods), as can be seen in Fig.~\ref{fig8}. As the value of $\epsilon$ increases, the absolute value of $V_{4S}$ also grows (the fourth tetrahedral site gets more attractive). The threshold value for desorbing the CNT corresponds to a $V_{4S}$ value close to $-k_B T$, the hydrophobicity limit already indicated. But what is more significant is that the threshold for complete filling occurs at a value of around $V_{4S}=-6kJ/mol$, the hydrophilicity value already indicated, signed by the $V_{4S}$ value of the HDA-like molecules in bulk water. This saturation threshold is far lower than the value corresponding to a linear hydrogen bond (which falls below -20kJ/mol) and shows that it suffices to provide a HDA-like environment in order to ensure full wetting, hence pointing again to the fact of the validity of the two-state picture of water for contexts that lie beyond bulk conditions.

As a final remark, it is interesting to note that a recent historical review on the structure of water indicated that while the two-liquid description would be relevant to the behavior of supercooled water, it is not clear that it would also be significant to our understanding of water and its interactions in real-life situations\cite{finney_historical}. In this sense, our results for bulk water indeed showed the existence of two local molecular arrangements, with clearly different second molecular layer organizations, all along the liquid regime that covers from supercooling to room temperature. Furthermore, our study of the interactions of water with different systems under ambient conditions actually revealed the central role that the two-liquid scenario might play within the hydration and nanoconfinement realms in ruling hydrophobicity and wetting.

\section{Conclusions}
In this work we presented a complete study of the $V_{4S}$ indicator for water, a generic structural index based on water's tetrahedrally-directional energetic interactions. This parameter has been specifically devised to be applied beyond bulk conditions, thus being suitable for hydration and nanoconfinement settings where previous indicators would produce artifacts. We have shown that such index enables to elucidate the existence of a fine-tuned interplay between local structure and energetics in bulk water that allows it to establish an extended hydrogen bond network while compensating for uncoordinated sites. This molecular mechanism underlies water’s two-liquid scenario, thus providing a detailed molecular rationale for this relevant description. Moreover, we showed that it is also operative at hydration and nanoconfinement conditions, thus extending the validity of the two-liquid scenario to such contexts. In so doing, this approach enabled to define conditions for wettability, thus providing an accurate measure of hydrophobicity and a reliable predictor of water filling and drying transitions. These goals had been out of reach in the absence of a proper molecular rationalization of the wetting process. As preliminary applications, we have applied the $V_{4S}$ index to graphene-like systems (graphene sheets, parallel graphene plates and carbon nanotubes) to explain their purported anomalous hydrophobicity and wetting. These results might hold the promise of opening new roads in contexts of paramount relevance in which water plays an active determining role, like biophysics and materials science.

\section{Methods}

In this work, we studied systems of bulk water and water in contact with self-assembled monolayers (SAMs), graphene sheets and narrow single-walled carbon nanotubes (CNT) \cite{v4s,review_Garde,graphene,cavities,Fluid-Phase-Equilibria,HummerCNT}. We performed molecular dynamic simulations using Groningen Machine for Chemical Simulations (GROMACS) package version 2022.4. \cite{gromacs}, employing the GAFF force field for all the systems we studied. We constrained bonds with the LINear Constraint Solver (LINCS) algorithm, and we used the Particle Mesh Ewald (PME) method to evaluate the long range electrostatics. We employed a Bussi--Donadio--Parrinello thermostat\cite{thermostat}(0.1 ps time constant) and a Parrinello--Rahman barostat (2.0 ps time constant) at 1 bar as reference pressure. In all cases, we generated production runs with a time step of 2 fs for all the systems that we investigated after equilibrating for timescales much larger than the $\alpha$ relaxation (that is, the self-intermediate scattering function has decayed to $1/e$), and we built boxes of appropriate sizes with periodic boundary conditions and a cutoff of 0.9 nm for the short range forces.
The bulk water systems were simulated for 2 ns within cubic boxes containing 10000 water molecules at 300K for the SPC/E system and in the range from 200K to 300K for TIP4P/2005. All the non-bulk systems (that is the SAMs, the graphene layers and the CNT) were solvated with  TIP4P/2005 water molecules and simulated at a temperature of 300K. Also, in all the stabilizations and dynamics, the initial position of the SAMs, the graphene layers and the CNT was restrained (center of the box). The pressure was 1 bar for all systems in this work.
We built SAMs containing 522 alkane-thiol surfactant chains (261 sulfurs) following an already known method\cite{SAMS} which, in summary, consisted in two surfactant chains each comprising $C_{10}$ alkane chains with a head group at one end and the other one attached to the sulfur atoms (which were position-restrained to locations corresponding to those of the gold 111 lattice\cite{gold}). The head groups were functionalized with hydrophilic ($OH$) or hydrophobic ($CH_3$) groups\cite{v4s,review_Garde} (this two systems were called as $SAM-OH$ and $SAM-CH_3$ respectively). Afterwards, water was added in the x, y and z directions to finally obtain a box with dimensions of 20 nm x 20 nm x 10 nm giving a total of 9800 water molecules.
The graphene sheet consisted of a perfect honeycomb graphite-like sheet with 416 carbon atoms with terminations in hydrogen atoms. For the double layer setting, two of such sheets were placed parallel one to another with a separation between the plates of 7.0\AA\ and 6.5\AA. Both systems were solvated to finally obtain boxes with dimensions of 12 nm x 12 nm x 7 nm, each one with more than 31200 water molecules.
Finally a CNT 112-carbon nanotube 11\AA\ long and 7.4\AA\ diameter was formed and solvated in a cubic box of 7 nm.

\section{Acknowledments}

This work was possible thanks to public funding. The authors acknowledge support form CONICET, UNS and ANPCyT (PICT2017/3127).

\bibliography{bibliografia} %Archivo con datos de referencias
\bibliographystyle{jcp} %Estilo de bibliografía generado manualmente con $ latex makebst

\end{document}